\documentclass[11pt]{article}

\usepackage[T1]{fontenc}
\usepackage{lmodern}
\usepackage{amsmath,amssymb,amsthm,mathtools}
\usepackage[margin=1in]{geometry}
\usepackage{microtype}
\usepackage[numbers,sort&compress]{natbib}
\usepackage[hidelinks]{hyperref}
\providecommand{\doi}[1]{\href{https://doi.org/#1}{DOI: \nolinkurl{#1}}}
\usepackage{booktabs}
\usepackage{array}
\usepackage{enumitem}

\numberwithin{equation}{section}

\theoremstyle{plain}
\newtheorem{theorem}{Theorem}[section]
\newtheorem{lemma}[theorem]{Lemma}
\newtheorem{corollary}[theorem]{Corollary}

\theoremstyle{definition}
\newtheorem{definition}[theorem]{Definition}

\newcommand{\Aut}{\operatorname{Aut}}
\newcommand{\Stab}{\operatorname{Stab}}
\newcommand{\Trans}{\operatorname{Trans}}
\newcommand{\UT}{\operatorname{UT}}

\newcommand{\Z}{\mathbb{Z}}
\newcommand{\Iso}{\operatorname{Iso}}
\newcommand{\Isot}{\operatorname{Isot}}

\title{On Applications of Polynomial-Time Group Isomorphism Algorithm}
\author{
Wenkang Zhang\textsuperscript{*$\dagger$},
Zhinan Wang\textsuperscript{*$\dagger$},
Jincheng Zhuang\textsuperscript{*$\dagger$}\\[4pt]
\textsuperscript{*}School of Cyber Science and Technology,\\
Shandong University, Qingdao 266237, China\\
\textsuperscript{$\dagger$}State Key Laboratory of Cryptography and Digital Economy Security,\\
Shandong University, Qingdao 266237, China\\
\{202517247, 202316997\}@mail.sdu.edu.cn,\\
jzhuang@sdu.edu.cn\\
}

\date{November 13, 2026}

\begin{document}

\maketitle

\begin{abstract}
The group isomorphism problem is a fundamental problem in computational group theory and is closely related to many other problems. Recently, Qiao and Zhang proposed a polynomial-time algorithm for finite group isomorphism in the multiplication-table model that computes the coset of all isomorphisms. Based on the algorithm of Qiao and Zhang, we study applications to finite associative ring isomorphism, bilinear map isotopism over finite commutative rings and free three-tensor isomorphism over finite commutative rings. First, we obtain a polynomial-time algorithm that returns the coset of all isomorphisms for finite associative rings. Second, we extend efficient isotopism and isomorphism algorithms for bilinear maps and free three-tensors from fields to general finite commutative rings. 
\end{abstract}

\noindent\textbf{Keywords:} Group isomorphism; Finite rings; Bilinear maps; Three-tensor isomorphism.

\medskip
\noindent\textbf{MSC 2020:} 68Q25, 20D15, 16Z05.

\section{Introduction}

The finite group isomorphism problem asks whether two groups given by multiplication tables are isomorphic. It is a fundamental problem in computational group theory and is closely related to other important problems in ring theory, bilinear algebra, and three-tensor isomorphism. In a recent breakthrough, Qiao and Zhang \cite{QZ26} gave an algorithm that computes the coset of all isomorphisms between two finite groups of order $N$ given by multiplication tables in time $N^{O(1)}$.

Prior to this breakthrough, polynomial-time algorithms were known for several classes of groups, including groups with no nontrivial abelian normal subgroups \cite{BCQ12} and certain families of solvable groups \cite{LeGall09,BQ12,BMW17}, while the general case remained unresolved. For $p$-groups of order $N$, class $2$, and exponent $p$ for odd primes $p$, Sun obtained an $N^{O((\log N)^{5/6})}$-time algorithm \cite{Sun23}. Ivanyos et al. improved this bound to $N^{\widetilde O(\sqrt{\log N})}$ and extended it to $p$-groups of Frattini class $2$ for odd $p$ \cite{IMQSZ24}. For arbitrary groups of order $N$ given by multiplication tables, the classical generator-enumeration algorithm runs in time $N^{\log_p N+O(1)}$, where $p$ is the smallest prime divisor of $N$ \cite{Miller78}. Subsequent improvements reduced the coefficient of $\log_p N$ to $1/2$ via bidirectional collision detection \cite{Rosenbaum13} and further to $1/4$ \cite[Section~2.2]{LGR16}.

Ring isomorphism is another classical problem in computational algebra. A finite ring $R$ with identity, not necessarily commutative, can be recovered from the unitriangular group $\UT_3(R)$ with two distinguished elements, as established by Mal'cev \cite{Malcev65} and further studied by Belegradek \cite{Belegradek93}. For rings given in a compact basis representation, Kayal and Saxena \cite{KS05,KS06} established complexity connections with graph isomorphism and integer factorization through ring isomorphism and automorphism problems. For rings given by operation tables, the natural generator-enumeration algorithm runs in time $N^{\log_p N+O(1)}$, where $p$ is the smallest prime divisor of $N$, by enumerating the possible images of an additive generating set of size at most $\log_p N$. Rosenbaum \cite[Section~4]{Rosenbaum13} subsequently improved this bound to $N^{(1/2)\log_p N+O(1)}$.

Bilinear map isotopism and tensor isomorphism are closely related problems in computational algebra. These problems have applications in post-quantum cryptography. For example, Ji, Qiao, Song, and Yun \cite{JQSY19} proposed tensor-based group actions as candidates for cryptographic constructions, while Tang et al. \cite{Tang22} constructed digital signature schemes based on the equivalence of alternating trilinear forms. Over finite fields, bilinear map isotopism is equivalent to three-tensor isomorphism. Qiao and Zhang \cite{QZ26} gave a $q^{O(n)}$-time algorithm for three-tensor isomorphism over the finite field $\mathbb{F}_q$. 
Grochow, Qiao, Stange, and Sun \cite{GQSS25} developed a complexity-theoretic framework for three-tensor isomorphism over both general and free modules and established completeness results for several natural algebraic problems.

\noindent\textbf{Our contribution.} Building on the algorithmic breakthrough of Qiao and Zhang \cite{QZ26}, we obtain the following results. Table~\ref{tab:1} summarizes our results and compares them with previous bounds.

\begin{table}[htb]
\centering
\footnotesize
\hyphenpenalty=10000
\exhyphenpenalty=10000
\setlength{\tabcolsep}{4pt}
\begin{tabular}{@{}
>{\centering\arraybackslash}m{0.20\textwidth}
>{\centering\arraybackslash}m{0.22\textwidth}
>{\centering\arraybackslash}m{0.30\textwidth}
>{\centering\arraybackslash}m{0.18\textwidth}
@{}}
\toprule
Problem & Parameters & Previous results & This paper \\
\midrule

Ring isomorphism &
Associative rings with identity; order $N$; two operation tables &
$N^{(1/2)\log_p N+O(1)}$ \cite{Rosenbaum13} &
$N^{O(1)}$ \\
\midrule

Bilinear-map isotopism &
Bilinear maps over a commutative ring; $M=|R|^3|U||V||W|$ &
No uniform polynomial-order bound known; quasi-polynomial for several cases \cite{GQSS25} &
$M^{O(1)}$ \\
\midrule

Free three-tensor isomorphism over a general commutative ring &
Free three-tensors over a commutative ring $R$; dimensions $a,b,c$ &
No uniform polynomial-order bound known for general commutative rings; quasi-polynomial for several cases \cite{GQSS25} &
$|R|^{O(a+b+c+1)}$ \\

\bottomrule
\end{tabular}

\caption{Comparison between previous and our results}
\label{tab:1}
\end{table}
\smallskip

\begin{enumerate}[label=(\arabic*)]
    \item Ring isomorphism. We give a polynomial-time algorithm for computing an isomorphism together with generators of the automorphism group for \textit{arbitrary finite associative rings}. Our approach uses the classical unitriangular construction with two named elements \cite{Belegradek93} to reduce ring isomorphism to marked-group isomorphism, and then recovers the ring isomorphisms from the resulting group isomorphisms.
    \item Bilinear map isotopism and free three-tensor isomorphism over finite commutative rings. We give a polynomial-time algorithm for computing an isotopism and generators of the autotopism group for bilinear maps over \textit{arbitrary finite commutative rings}. Our approach encodes a bilinear map into a triangular ring, where the diagonal idempotents and scalar ring recover the three module actions. As a corollary, we obtain an algorithm for free three-tensor isomorphism over finite commutative rings.
\end{enumerate}

The remainder of this paper is organized as follows. Section~\ref{sec:prelim} specifies the models, definitions, and imported algorithms. Sections~\ref{sec:rings}--\ref{sec:bilinear} present our algorithms and establish the main results. Finally, Section~\ref{sec:conclusion} concludes the paper.

\section{Preliminaries}
\label{sec:prelim}

\subsection{Notation and Definitions}

\medskip
\noindent\emph{\textbf{Notation}.}

Direct sums and tensor products are denoted by
$\oplus$ and $\otimes$, respectively, with $\otimes_R$
indicating a tensor product over $R$.
For algebraic structures $X,Y$, $\Aut(X)$ denotes
its automorphism group, and $\Iso(X,Y)$ denotes
the set of all isomorphisms from $X$ to $Y$.
For algebraic structures $b,b'$,
we write $\Isot(b,b')$ for the set of all isotopisms
from $b$ to $b'$.

For complexity bounds, $N$ denotes the order
of the algebraic structure under consideration. $L_{\mathrm{bit}}$ denotes the total bit length of the relevant problem input.
The symbols $O(\cdot)$ and $\mathrm{poly}(\cdot)$
have their usual asymptotic meanings.

\medskip

\begin{definition}[Central elements]
An element of a group is central if it commutes with every element.
\end{definition}

\begin{definition}[Bilinear maps and isotopism]
Let $R$ be a commutative ring with identity, and let $U,V,W$
be finite $R$-modules. A map $b\colon U\times V\to W$ is
$R$-bilinear if it is additive in each argument and
$b(ru,v)=r\,b(u,v)=b(u,rv)$ for all $r\in R$, $u\in U,$ and $v\in V$.
An isotopism from $b$ to another bilinear map
$b'\colon U'\times V'\to W'$ is a triple
\((A,B,C)\) of \(R\)-module isomorphisms
\[
A\colon U\to U',\qquad
B\colon V\to V',\qquad
C\colon W\to W',
\]
with
\[
C(b(u,v))=b'(Au,Bv).
\]

\end{definition}

\begin{definition}[Three-tensors]
Let $R$ be a commutative ring with identity.
A three-tensor $T$ over $R$ is an element of
$U\otimes_R V\otimes_R W$ for finite $R$-modules $U,V,W$.
It is free if $U,V,W$ are free; a free three-tensor
is thus an element of
$R^a\otimes_R R^b\otimes_R R^c$ for some $a,b,c\ge1$.
An isomorphism from $T$ to $T'$ is a triple $(A,B,C)$
of $R$-module isomorphisms with
\[
T'=(A\otimes B\otimes C)T;
\]
in the free case these are automorphisms of $R^a,R^b,R^c$.
We write $\Iso(T,T')$ for the set of all such isomorphisms.
\end{definition}

\begin{definition}[Transporters and cosets]
We now fix the output convention used throughout the paper.
If a group $H$ acts on a set $X$ on the left, define
\[
\Trans_H(x,y)=\{h\in H:h\cdot x=y\},
\qquad
\Stab_H(x)=\{h\in H:h\cdot x=x\}.
\]
A nonempty transporter is the coset
$h_0\Stab_H(x)=\{h_0 s:s\in\Stab_H(x)\}$.
We say that an algorithm computes this transporter if it
outputs $h_0$ and a generating set of $\Stab_H(x)$,
and outputs $\varnothing$ otherwise.

For algebraic structures $X$ and $Y$, if $X\cong Y$,
the set $\Iso(X,Y)$ of all isomorphisms from $X$ to $Y$
is the coset
$\varphi\circ\Aut(X)=
\{\varphi\circ\alpha:\alpha\in\Aut(X)\}$
for any $\varphi\in\Iso(X,Y)$.
We say that an algorithm computes the coset
$\Iso(X,Y)$ if it outputs one isomorphism
$\varphi\in\Iso(X,Y)$ and a generating set
of $\Aut(X)$ when $X\cong Y$,
and outputs $\varnothing$ otherwise.

The same convention applies to isotopisms:
if $b$ and $b'$ are isotopic, the set
$\Isot(b,b')$ of all isotopisms from $b$ to $b'$
is a coset of the autotopism group of $b$,
and we say that an algorithm computes this coset
when it outputs one isotopism and generators
of the autotopism group.
The autotopism group of $b$ is its stabilizer
under independent $R$-module automorphisms
of $U$, $V$, and $W$.
\end{definition}

\subsection{Group Isomorphism and Algorithmic Preliminaries}

We use the following result from Qiao and Zhang \cite{QZ26}. 

\begin{theorem}[Group isomorphism {\cite[Theorem~9.1]{QZ26}}]
\label{thm:gpiso}
For two multiplication-table groups $G$ and $H$ of order $N$, the coset $\Iso(G,H)$ can be computed in time $N^{O(1)}$.
\end{theorem}

The algorithm of Qiao and Zhang returns the coset of all isomorphisms between two structures, represented by one representative and generators of the corresponding automorphism group. To pass from groups to rings and bilinear maps, we must restrict such a coset by requiring selected source elements to be sent to selected target elements (marks). The following lemma shows that these restrictions can be imposed directly on the coset representation, in time polynomial in the number of marks and the permutation degree, without enumerating the coset. It is the basic operation behind the marked constructions in Sections~\ref{sec:rings} and~\ref{sec:bilinear}; for example, it imposes the two marks in $\UT_3(R)$ and the diagonal idempotent and scalar marks in the triangular-ring construction.

\begin{lemma}[Imposing individual marks]
\label{lem:marks}
Suppose a representative and permutation-group generators describing the coset of all isomorphisms between two structures with $M$ elements are given. A list of $s$ requirements sending specified source elements to specified target elements can be imposed in time polynomial in $M$, $s$, and the description length of the set.
\end{lemma}

\begin{proof}
Maintain a nonempty current set \(\varphi K\), where \(K\) fixes all source
marks processed so far. For the next constraint \(\varphi k(a)=b\), compute
the image set
\[
K(a)=\{k(a):k\in K\}.
\]
If \(\varphi^{-1}(b)\notin K(a)\), the constrained set is empty. Otherwise
choose \(k_0\in K\) with \(k_0(a)=\varphi^{-1}(b)\), replace \(\varphi\) by
\(\varphi k_0\), and replace \(K\) by the subgroup of \(K\) fixing \(a\),
\[
K_a=\{k\in K:k(a)=a\}.
\]
Then \(\varphi k_0K_a\) is exactly the subset of \(\varphi K\) satisfying
\(\varphi k(a)=b\): every such \(k\) differs from \(k_0\) by an element
fixing \(a\). Standard permutation-group algorithms compute these image sets
and fixing subgroups in polynomial time in the permutation degree
\cite{Seress03}. The procedure does not enumerate all images of the tuple of
marks.

\end{proof}

\section{Finite associative ring isomorphism}
\label{sec:rings}

We reduce ring isomorphism to isomorphism of two-marked groups via the
unitriangular construction, compute the coset of all marked group
isomorphisms, and recover the coset of all ring isomorphisms. Each ring below
is given by its addition and multiplication tables.

\subsection{Reduction from ring isomorphism to group isomorphism}

The unitriangular interpretation below goes back to Mal'cev \cite{Malcev65} and Belegradek \cite{Belegradek93}. Let $R$ be a finite associative ring with identity. On the set $R^3$ define
\begin{equation}
\label{eq:group-law}
(a,b,c)(a',b',c')=(a+a',\,b+b',\,c+c'+ab').
\end{equation}
This is the unitriangular group $H(R)=\UT_3(R)$. Associativity follows from distributivity: the third coordinate of either parenthesization contains the terms $ab'+ab''+a'b''$ in addition to the three central coordinates. Its identity is $(0,0,0)$ and
\[
(a,b,c)^{-1}=(-a,-b,-c+ab).
\]
Mark the two elements $x=(1,0,0)$ and $y=(0,1,0)$. We give the recovery argument to keep the output conversion explicit.

\begin{lemma}[Recovering the ring]
\label{lem:recover}
A group isomorphism $H(R)\to H(S)$ sending $x$ to $x'$ and $y$ to $y'$ induces a unital ring isomorphism $R\to S$. Every ring isomorphism arises in this way.
\end{lemma}

\begin{proof}
Use the convention $[g,h]=ghg^{-1}h^{-1}$. Direct calculation gives
\begin{equation}
\label{eq:commutator}
[(a,b,c),(a',b',c')]=(0,0,ab'-a'b).
\end{equation}
Since the rings have identities, an element is central exactly when $a=b=0$. Hence the center is $Z=\{(0,0,c):c\in R\}$. The centralizers of the two marks are
\[
C_{H(R)}(x)=\{(a,0,c)\},
\qquad
C_{H(R)}(y)=\{(0,b,c)\}.
\]
Thus a marked group isomorphism induces additive bijections $A,B,C\colon R\to S$ on $C(x)/Z$, $C(y)/Z$, and $Z$, respectively. Preservation of \eqref{eq:commutator} gives
\begin{equation}
\label{eq:C-product}
C(ab)=A(a)B(b).
\end{equation}
The marks imply $A(1)=B(1)=1$. Setting $b=1$ and then $a=1$ in \eqref{eq:C-product} gives $C=A=B$. Consequently $C$ preserves addition, multiplication, and the identity. Conversely, a ring isomorphism $f\colon R\to S$ defines the marked group isomorphism $(a,b,c)\mapsto(f(a),f(b),f(c))$ by \eqref{eq:group-law}.
\end{proof}

The induced map from marked group automorphisms to ring automorphisms is
surjective but need not be injective: central additive changes can disappear
under the projection. Since a surjective homomorphism maps generators to
generators, the images of the computed generators generate the ring
automorphism group; injectivity is not required.

\subsection{The ring isomorphism algorithm}

\begin{theorem}[Finite associative ring isomorphism]
\label{thm:ringiso}
Given addition and multiplication tables of two finite associative rings $R$ and $S$ with identity and common order $N$, one can compute the coset $\Iso(R,S)$ of all ring isomorphisms between them in time $N^{O(1)}$.
\end{theorem}

\begin{proof}
Construct $H(R)$ and $H(S)$ and their multiplication tables. Each group has $N^3$ elements and its table has $N^6$ entries. Apply Theorem~\ref{thm:gpiso}. If its output is nonempty, impose the two mark constraints with Lemma~\ref{lem:marks}. By Lemma~\ref{lem:recover} the resulting set is nonempty exactly when the rings are isomorphic.

Restrict a representative and the marked source stabilizer generators to the center. Through the explicit identifications of the centers with the additive rings, these restrictions give a ring-isomorphism representative and generators of $\Aut(R)$. Surjectivity in Lemma~\ref{lem:recover} proves that this output is the coset $\Iso(R,S)$. All table construction, point constraints, and restrictions take time polynomial in $N$. Unequal ring orders are rejected before this procedure. The one-element ring is handled directly.
\end{proof}

\begin{corollary}[Rings without identity]
\label{cor:nonunital}
Let $R$ and $S$ be finite associative rings without identity, given by addition
and multiplication tables, and let $m$ and $m'$ be their additive exponents,
i.e.\ the least positive integers with $mR=0$ and $m'S=0$. If $|R|\ne |S|$ or
$m\ne m'$, reject. Otherwise the coset $\Iso(R,S)$ of all ring isomorphisms
between $R$ and $S$ can be computed in time polynomial in $|R|$ and $|S|$.
\end{corollary}

\begin{proof}
The construction below is the truncated Dorroh extension \cite{Dorroh32}. Assume $m=m'$. Define
\[
A_R=(\Z/m^2\Z)\oplus R,\qquad
(a,r)(b,s)=(ab,\,as+br+rs),
\]
where integers act on $R$ through its additive group. This is an associative
unital ring with identity $(1,0)$. Put $t_R=(m,0)$. Then $t_R=m(1,0)$ is
central.

Set
\[
J_R=\{x\in A_R:xt_R=t_Rx=0\},\qquad I_R=\Z t_R.
\]
Since $mR=0$, a direct calculation gives
\[
J_R=m(\Z/m^2\Z)\oplus R,\qquad
I_R=m(\Z/m^2\Z)\oplus0.
\]
The map $J_R\to R$, $(ma,r)\mapsto r$, is a surjective ring homomorphism with
kernel $I_R$; hence $J_R/I_R\cong R$.

Define $A_S,t_S,J_S,I_S$ analogously. Every unital ring isomorphism
$A_R\to A_S$ sends $t_R$ to $t_S$, hence maps $J_R$ to $J_S$ and $I_R$ to
$I_S$, and therefore induces a ring isomorphism
$R\cong J_R/I_R\to J_S/I_S\cong S$. Conversely, every ring isomorphism
$R\to S$ extends to a unital ring isomorphism $A_R\to A_S$ by fixing the first
coordinate. Apply Theorem~\ref{thm:ringiso} to $A_R$ and $A_S$, and restrict
the resulting isomorphisms to $J_R/I_R$. The order of $A_R$ is
$m^2|R|\le |R|^3$, since $m$ divides $|R|$; all table operations are therefore
polynomial in $|R|$. The one-element ring is handled separately.
\end{proof}

\section{Bilinear map isotopism and free three-tensor isomorphism over a finite commutative ring}
\label{sec:bilinear}

We apply the ring isomorphism algorithm of Section~\ref{sec:rings} to the triangular ring constructed from a bilinear map that is represented by its full value table. We encode isotopism in a marked triangular ring, obtain the coset of all isotopisms, and dualize to free three-tensors. 

Each bilinear map $b\colon U\times V\to W$ below is given by its full table $(u,v)\mapsto b(u,v)$ for all $(u,v)\in U\times V$, and similarly for $b'$. Each $R$-module $U$ below is given by its addition table and the scalar-action table $(r,u)\mapsto ru$ for all $r\in R$ and $u\in U$.

\subsection{Encoding a bilinear map in a ring}
We use the standard triangular matrix ring construction \cite[Chapter~2]{AndersonFuller92}. Fix a finite commutative ring $R$ with identity. Let $U,V,W$ and $U',V',W'$ be finite $R$-modules, and let
\[
b\colon U\times V\to W,
\qquad
b'\colon U'\times V'\to W'
\]
be bilinear maps. We characterize an isotopism from \(b\) to \(b'\) by a triple
\((A,B,C)\) of \(R\)-module isomorphisms:
\(A\colon U\to U'\), \(B\colon V\to V'\), \(C\colon W\to W'\)
satisfying
\begin{equation}
\label{eq:isotopism}
C(b(u,v))=b'(Au,Bv).
\end{equation}

Define the additive group $S_b=R^3\oplus U\oplus V\oplus W$. An element is written $(r_1,r_2,r_3,u,v,w)$, with coordinate-wise addition
\[
(r_1,r_2,r_3,u,v,w)+(s_1,s_2,s_3,u',v',w')
=(r_1+s_1,\,r_2+s_2,\,r_3+s_3,\,u+u',\,v+v',\,w+w')
\]
and zero $(0,0,0,0,0,0)$. Define multiplication by
\begin{equation}
\label{eq:ring-product}
\begin{aligned}
&(r_1,r_2,r_3,u,v,w)(s_1,s_2,s_3,u',v',w')\\
&\quad=(r_1s_1,\,r_2s_2,\,r_3s_3,\,r_1u'+s_2u,\,r_2v'+s_3v,\,r_1w'+s_3w+b(u,v')).
\end{aligned}
\end{equation}

Thus \(S_b\) is the upper triangular matrix ring
\[
\begin{pmatrix}
R&U&W\\
0&R&V\\
0&0&R
\end{pmatrix},
\]
with the \((1,3)\)-entry multiplied via
\(b\colon U\times V\to W\); its unit is \((1,1,1,0,0,0)\).

For \(i=1,2,3\), let \(e_i\in S_b\) be the element with \(1\) in the
\(i\)-th diagonal coordinate and \(0\) elsewhere, and for \(r\in R\) set
\[
\Delta(r)=(r,r,r,0,0,0).
\]
We call \(e_1,e_2,e_3\) the diagonal idempotents and the elements
\(\Delta(r)\) the scalar diagonal elements.

A marked ring is a ring together with a distinguished set of elements.
A marked ring isomorphism is a ring isomorphism that sends each
distinguished element to the corresponding distinguished element.

\begin{lemma}[Marked triangular-ring correspondence]
\label{lem:triangular}
Let $R$ be a finite commutative ring with identity. With coordinate-wise addition, formula~\eqref{eq:ring-product} defines an associative ring. If its three diagonal idempotents and the scalar diagonal elements are individually marked, the coset of all isomorphisms between two such marked rings is naturally identified with the coset of all isotopisms between the corresponding bilinear maps.
\end{lemma}

\begin{proof}
Distributivity follows from the additivity of $b$ and the module axioms; the only nontrivial point is associativity. Associativity of the diagonal coordinates and the first two off-diagonal coordinates follows from the ring and module axioms. In the final coordinate, the terms containing $b$ agree because
\[
b(ru,v)=r\,b(u,v)=b(u,rv)
\]
and because $b$ is additive in both arguments. In a triple product, for example, $b(r_1u'+s_2u,v'')=r_1b(u',v'')+b(u,s_2v'')$, and multiplying $b(u,v')$ by the final diagonal scalar agrees with multiplying $v'$ by that scalar first. These are precisely the additional identities needed in the final coordinate. Hence \eqref{eq:ring-product} is associative.

Recall that $e_1,e_2,e_3$ are the diagonal idempotents and $\Delta(r)=(r,r,r,0,0,0)$. Impose $e_i\mapsto e_i'$ and $\Delta(r)\mapsto\Delta'(r)$ for all $r\in R$, where \(e_i'\) and \(\Delta'(r)\) denote the corresponding elements of \(S_{b'}\). The corners
\[
e_1S_be_2=U,
\qquad
e_2S_be_3=V,
\qquad
e_1S_be_3=W
\]
are then preserved. The diagonal corners are fixed pointwise, since their elements are $e_i\Delta(r)$. Restrictions to the three off-diagonal corners give additive bijections $A,B,C$. Fixing $\Delta(R)$ makes them $R$-linear. Multiplying elements in the first two corners recovers $b$, so preservation of multiplication is exactly \eqref{eq:isotopism}.

Conversely, an isotopism extends to a ring isomorphism by fixing the three diagonal coordinates and applying $A,B,C$ to the off-diagonal coordinates. Formula~\eqref{eq:ring-product} verifies multiplicativity. The extension is unique because the six additive corners determine every element. Thus the correspondence is a bijection between the two sets of isomorphisms. It is compatible with composition, hence induces an isomorphism between the corresponding automorphism groups.
\end{proof}

\subsection{General modules and free three-tensors}

\begin{theorem}[Bilinear maps over a finite commutative ring]
\label{thm:bilinear}
The isotopism coset $\Isot(b,b')$ can be computed in time $M^{O(1)}\mathrm{poly}(L_{\mathrm{bit}})$, where, after checking corresponding cardinalities,
\[
M=|R|^3|U||V||W|.
\]
\end{theorem}

\begin{proof}
Reject if corresponding module cardinalities differ. Build the two rings in Lemma~\ref{lem:triangular}; they both have order $M$. Apply Theorem~\ref{thm:ringiso} and then impose the diagonal idempotent and scalar marks using Lemma~\ref{lem:marks}. There are at most $|R|+3$ marks, and $|R|\le M$. Finally restrict the representative and generators to the off-diagonal corners. Lemma~\ref{lem:triangular} proves both correctness and that the output is the isotopism coset $\Isot(b,b')$. The construction and all orbit-stabilizer operations are polynomial in $M$ and the explicit input length.
\end{proof}

\begin{corollary}[Free three-tensors over a finite commutative ring]
\label{cor:three-tensors}
Let \(R\) be a finite commutative ring with identity. For two free three-tensors $T,T'\in R^a\otimes_R R^b\otimes_R R^c$, under independent general linear changes of basis in the three factors, there is an algorithm that returns the coset $\Iso(T,T')$ of all isomorphisms between them in time
\[
|R|^{O(a+b+c+1)}\mathrm{poly}(L_{\mathrm{bit}}).
\]
\end{corollary}

\begin{proof}
A tensor in three finite free modules is equivalently a bilinear map from the duals of the first two modules to the third. Dualization converts the first two tensor actions into inverse-transpose actions, which still range over the full general linear groups. Under this correspondence, tensor isomorphisms and automorphisms are identified with bilinear-map isotopisms and autotopisms via the appropriate dual transformations. Applying Theorem~\ref{thm:bilinear} with module cardinalities $|R|^a$, $|R|^b$, and $|R|^c$, we can therefore recover a representative and generators of the tensor isomorphism coset $\mathrm{Iso}(T, T')$ within the stated time bound.

\end{proof}

\section{Conclusion}
\label{sec:conclusion}

In this work, we develop algorithms for several algebraic isomorphism problems by combining Qiao and Zhang's result \cite{QZ26}. We give a polynomial-time algorithm for computing isomorphisms and automorphism-group generators of arbitrary finite associative rings. We also obtain efficient algorithms for bilinear map isotopism over arbitrary finite commutative rings and free three-tensor isomorphism over arbitrary finite commutative rings.

Our main approach is to develop reductions from these algebraic problems to problems that can be solved using Qiao and Zhang's technique. Specifically, we reduce ring isomorphism to marked-group isomorphism through the unitriangular construction. We then reduce bilinear map isotopism to marked-ring isomorphism through a triangular ring construction. Finally, free three-tensor isomorphism is handled via its equivalence to bilinear map isotopism over finite free modules. These reductions preserve the information needed to recover not only individual isomorphisms but also generators of the corresponding automorphism groups. Future work may explore similar reductions for broader classes of algebraic structures and more compact input representations.

\section*{AI Disclosure}

We use large language models in preparing this paper,
including checking the algorithm, revising parts of the text. The authors take full responsibility for the content.

\begingroup
\renewcommand{\bibfont}{\normalfont\raggedright}
\bibliographystyle{unsrtnat}
\bibliography{references}
\endgroup

\end{document}